\documentclass[aps,prl,twocolumn,superscriptaddress,10pt]{revtex4-2}
\usepackage{graphicx} 
\usepackage{amsmath,amssymb,latexsym,epsfig,graphics,epsf,bm}
\usepackage{graphics,tabularx}
\newcolumntype{Y}{>{\hsize=0.9\hsize\centering\arraybackslash}X}
\newcolumntype{s}{>{\hsize=1.5\hsize\arraybackslash}X}

\newcommand{\de}{\Delta\varepsilon}
\newcommand{\es}{\varepsilon_\mathrm{s}}
\newcommand{\ei}{\varepsilon_\infty}
\newcommand{\ez}{\varepsilon_0}

\newcommand{\edp}{\varepsilon^{\prime\prime}(\nu)}
\newcommand{\gk}{g_\mathrm{K}}
\newcommand{\fig}[1]{Fig.~\ref{#1}}
\newcommand{\Eq}[1]{Eq.~(\ref{#1})}

\usepackage[symbol*]{footmisc}

\begin{document}

\title{Dipolar order controls dielectric response of glass-forming liquids} 

\author{Till Böhmer}\affiliation{Institute for Condensed Matter Physics, Technical University of Darmstadt, D-64289 Darmstadt,Germany} 
\author{Florian Pabst}\affiliation{Institute for Condensed Matter Physics, Technical University of Darmstadt, D-64289 Darmstadt,Germany}\affiliation{SISSA — Scuola Internazionale Superiore di Studi Avanzati, 34136 Trieste, Italy} 
\author{Jan P. Gabriel}\affiliation{{\it Glass and Time}, IMFUFA, Department of Science and Environment, Roskilde University, P. O. Box 260, DK-4000 Roskilde, Denmark} 
\author{Thomas Blochowicz}\affiliation{Institute for Condensed Matter Physics, Technical University of Darmstadt, D-64289 Darmstadt,Germany} 

\date{\today} 
 
\begin{abstract}

The dielectric response of liquids reflects both, reorientation of single molecular dipoles and collective modes, i.e., dipolar cross-correlations. A recent theory predicts the latter to produce an additional slow peak in the dielectric loss spectrum. Following this idea we argue that in supercooled liquids the high-frequency power law exponent of the dielectric loss $\beta$ should be correlated with the degree of dipolar order, i.e., the Kirkwood correlation factor $\gk$. This notion is confirmed for 25 supercooled liquids. While our findings support recent theoretical work the results are shown to violate the earlier Kivelson-Madden theory.

\end{abstract}

\maketitle


Supercooling a liquid below the melting temperature towards the glass transition while avoiding crystallization results in a dramatic slowing down of the molecular motions. Along with this slowdown it is observed that relaxation spectra of supercooled liquids significantly deviate from a Lorentzian shape, and are instead asymmetrically broadened with a high-frequency power law $\nu^{-\beta}$, $\beta\leq 1$, a phenomenon called \textit{relaxation stretching}. As both, the pronounced slowdown and the relaxation stretching, are universal manifestations of glassy dynamics, correlations between the two phenomena were suggested~\cite{Boehmer1993}, but a microscopic understanding is still lacking. Probably the most common picture to rationalize relaxation stretching is dynamical heterogeneity~\cite{Richert2002,Boehmer1998,Sillescu2002,Ediger2000}, meaning that the dynamics in different spatial regions contributes with a Lorentzian centered at different frequencies to the overall spectrum, which leads to the observed broadening. However, this is by far not the only explanation and theories exist which arrive at stretched structural relaxation without explicitly incorporation dynamical heterogeneity~\cite{Ngai2023,Goetze2009}. Additionally, it was found in computer simulations that -- although dynamic heterogeneity exists -- the relaxation curves are already intrinsically broadened when evaluated in small spatial regions~\cite{Berthier2021,Shang2019}.
Thus, clarifying the origin of the relaxation stretching can be considered of utmost importance for the understanding of the glass transition.


Over the years, relaxation stretching in molecular glass formers has been studied for a broad selection of supercooled liquids using dielectric spectroscopy, revealing a variety of high-frequency power law exponents $\beta$. Some years ago, Paluch et al.~\cite{Paluch2016} showed that for the dielectric loss, $\beta$ is correlated with the dielectric relaxation strength $\de=\es-\ei$, where $\es$ and $\ei$ are the zero and high frequency limits of the dielectric permittivity. Large $\de$ values are associated with higher values of $\beta$, i.e., highly polar liquids display less relaxation stretching than less polar ones, an observation which we refer to as \textit{Paluch correlation} hereafter.

Initially, the Paluch correlation was rationalized by arguing that strong dipole-dipole interactions increase the harmonicity of the inter-molecular potentials, which is thought to affect the distribution of relaxation times and, thus, the relaxation stretching~\cite{Paluch2016}. Following this line of ideas, the same degree of relaxation stretching observed for the dielectric loss should be observed in other experimental techniques. However, this conjecture is in stark contrast to recent experimental results: For instance in depolarized dynamic light-scattering (DDLS)~\cite{Pabst2021} and nuclear magnetic resonance~\cite{Koerber2020,Becher2021,Becher2021a}, both methods that probe molecular reorientation, a generic value $\beta\approx 0.5$ is observed for many supercooled liquids with very different polarities. Therefore, it can be concluded that the broad variety of dielectric relaxation shapes cannot primarily be a result of differences regarding the distribution of local relaxation times between more and less polar liquids.

Instead, recent results suggest that rather dynamic signatures of dipolar cross-correlations are responsible for the broad variety of relaxation stretching observed in dielectric experiments. This idea emerged from a careful comparison of dielectric and DDLS spectra ~\cite{Pabst2020,Pabst2021,Boehmer2022,Gabriel2020}, where it was found that the overall relaxation shape is a superposition of a rather generic distribution of self-correlation times with an additional slow and narrow process. The origin of the latter was traced back to dipolar cross-correlations. The suggested qualitative picture is that the self-part of dipolar dynamics in supercooled liquids has a generic high-frequency power law exponent of $\beta=0.5$ but is in many cases superimposed by a slow and narrow cross-correlation process effectively resulting in a steeper high-frequency power law exponent $\beta>0.5$. And indeed, the existence of the slow cross-correlation contribution in highly polar liquids was recently confirmed by computer simulations~\cite{Koperwas2022,Henot2023}. Nevertheless, the correct interpretation of the dielectric relaxation shape and the relevance of dipolar cross-correlations is still subject of controversial discussion\,~\cite{Ngai2021,Ngai2022,Moch2022,Moch2022a}. 

In an attempt to solve this important ongoing debate, we analyze in this letter the relation between the high-frequency power law exponent $\beta$ of the dielectric loss and the existence of dipolar cross-correlations for a broad variety of supercooled liquids. In this regard we consider the Kirkwood correlation factor $\gk$~\cite{Kirkwood1939} to quantify the strength of static dipolar cross-correlations. Simply speaking, $\gk>1$ indicates that orientations of neighbouring dipoles tend to show parallel alignment, while $\gk=1$ is attributed to the absence of dipolar order, i.e., no static cross-correlations exist.

\begin{figure}[t]
    \includegraphics[width=0.45\textwidth]{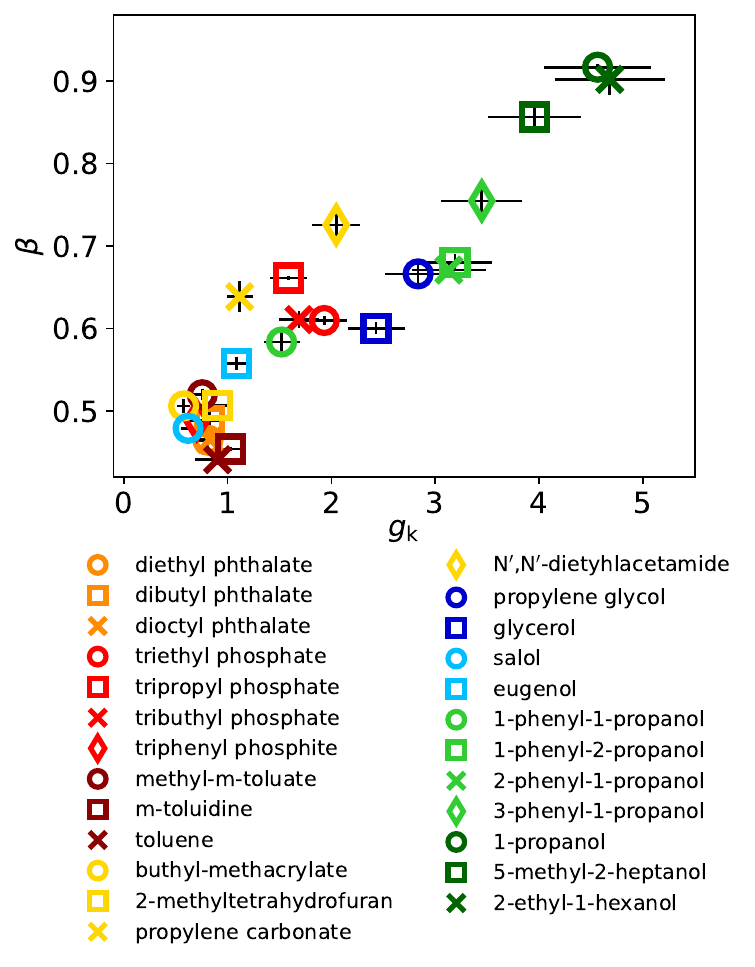}
    \caption{High-frequency power law exponent $\beta$ of the dielectric loss as function of the Kirkwood correlation factor $\gk$ for different supercooled liquids. Liquids with similar molecular structure share the same colors. Each data point represents the temperature-averaged values of $\gk$ and $\beta$; the temperature resolved version of the figure can be found in the SI. Errorbars consider the estimated uncertainty due to the uncertainty of parameters used in \Eq{equ:kb} (see SI for details), as well as the variation of $\gk$ and $\beta$ 
    as function of temperature.}
    \label{fig1}
\end{figure}

Our analysis considers dielectric data of 25 different supercooled liquids. As far as crystallization could be avoided, each liquid was studied at several temperatures in the deeply supercooled regime, i.e. $10^{-1}\,\mathrm{Hz}<\nu_\mathrm{peak}<10^5\,\mathrm{Hz}$. The reason for restricting the study to deeply supercooled liquids is that at these temperatures different processes have the maximum dynamic separation. Moreover, possible molecular interactions leading to cross-correlation effects are stronger with respect to the thermal energy. Consequently, the impact of cross-correlations is stronger at deeply supercooled temperatures compared to the mildly supercooled or liquid regime, where most likely additional effects control the shape of the relaxation spectrum.

The main result of our analysis is displayed in \fig{fig1}, revealing a strong correlation between the stretching parameter $\beta$ and the Kirkwood correlation factor $g_K$.

In order to determine $\beta$, we apply a model-free approach instead of using model functions, which are known to produce results depending on the exact procedure used. Instead, we follow Nielsen et al.\ \cite{Nielsen2009} and define $\beta$ as the logarithmic slope at the frequency of steepest ascent on the high-frequency flank of $\edp$:
\begin{equation}
    \beta = \min\left( \frac{\mathrm{d}\,\log\edp}{\mathrm{d}\,\log\nu}\right).
    \label{equ:beta}
\end{equation}
Full derivative spectra of several analyzed liquids are shown in the SI.

The values of $\gk$ were calculated as
\begin{equation}
    \gk = \frac{9k_\mathrm{B}\ez M T}{\rho N_\mathrm{A}\mu^2}\frac{(\es-\ei)(2\es+\ei)}{\es(\ei+2)^2},
    \label{equ:kb}
\end{equation}
where $T$ is temperature, $M$ molar mass, $\rho$ density and $\mu$ the gas-phase molecular dipole moment~\cite{Kirkwood1939}. Importantly, the molecular dipole moment $\mu$ needs to reflect the conformational states of the molecule in the liquid phase, thus $\mu$ should be determined via dilution experiments. Typically, no such data are available in the literature for more exotic molecules, like, e.g., most pharmaceuticals considered in Ref.~\cite{Paluch2016}. One of the reasons is that these molecules are quite large, flexible and consist of several functional groups that carry a substantial dipole moment. In this case, the total molecular dipole moment strongly depends on the conformational state of the molecule, making it impossible to assign one single value of $\mu$ in order to calculate $\gk$, let alone to determine $\mu$ in a dilution experiment. Thus, in the present study, we focus on smaller molecules, for which $\mu$ is readily available, which is the reason, why our considerations include a smaller number of substances compared to the work by Paluch et al.\ in Ref.~\cite{Paluch2016}. 

In brief, $\gk$ is calculated by extracting $\es$ from either our own dielectric measurements or in few cases from dielectric data received from other groups~(cf. Refs.~\cite{Casalini2003,Mandanici2005,Kudlik1999}). $\ei(T)=n(T)^2$ at optical frequencies was calculated from the refractive index $n(T)$, the temperature dependence of which was included via the Lorentz-Lorenz equation from $\rho(T)$, obtained by linear extrapolation of literature data to lower temperatures~\cite{Blazhnov2004}. Further details and all necessary parameters extracted from the literature~\cite{Kabachnik1963,Peshwe2009,Yaws2009,DeLorenzi1997,Kemppinen1956,Petkovic1973,Wang2009,Evans1930,Aroney1964,Mole1964,Sidgwick1934,Tommila1945,Barclay1951,Sulthana2020,Song2007,Wisniak2007,Faizullin2017,Francesconi2007,Wankhede2006,Mopsik1963,Weinstein2007,Ryabov2003,Copeland1951,Marti1930,Scaife1976,Kim1975,Romero2008,Shinomiya1989,Bhatia2013} are given in the SI.

It is important to note that both, $\gk$ and $\beta$ are temperature-dependent quantities~\cite{Kremer2002,Nielsen2009}. However, the temperature dependence of both is weak in the considered temperature range, and thus we plot in Fig.~\ref{fig1} one single temperature-averaged point per liquid. Temperature-resolved results are shown in the SI.

From \fig{fig1} it is immediately clear that higher values of $\beta$ are associated with larger $\gk$, i.e., higher dipolar order. Around $\gk\approx 1$ and $\beta\approx0.5$, a clustering of data points from different substances is observed. This agrees with several recent results from different experimental techniques identifying $\beta=0.5$ as the generic high-frequency power law exponent. In accordance with the ideas discussed above, the generic power law exponent is observed in dielectric spectroscopy as soon as dipolar cross-correlations are absent ($\gk\approx 1$).

At the other extreme are 1-propanol and 2-ethyl-1-hexanol, two monohydroxy alcohols for which we find $\gk\approx 5$ in agreement with their strong tendency to form supramolecular structures via hydrogen bonding. In the dielectric loss spectrum this is manifested by the appearance of an additional strong and slow Debye process that superimposes the $\alpha$-process. In contrast to most other classes of liquids, it is well-established for monohydroxy alcohols that dipolar cross-correlations due to formation of supra-molecular structures are the origin of the additional Debye process~\cite{Boehmer2014}. Thus, the Debye-process in monohydroxy alcohols is treated in the same way as the cross-correlation contribution in any other liquid.

In between both extremes, at $1<\gk<5$, we find a broad variety of different substances with increasingly large values of $\beta$. These liquids have very different chemical structures and can be sorted into different sub-categories, namely non-hydrogen bonding liquids with and without aromatic groups, hydroxy aromatics, polyhydric alcohols, as well as monohydroxy alcohols with and without aromatic groups.

\begin{figure}
\centering
    \includegraphics[width=0.45\textwidth]{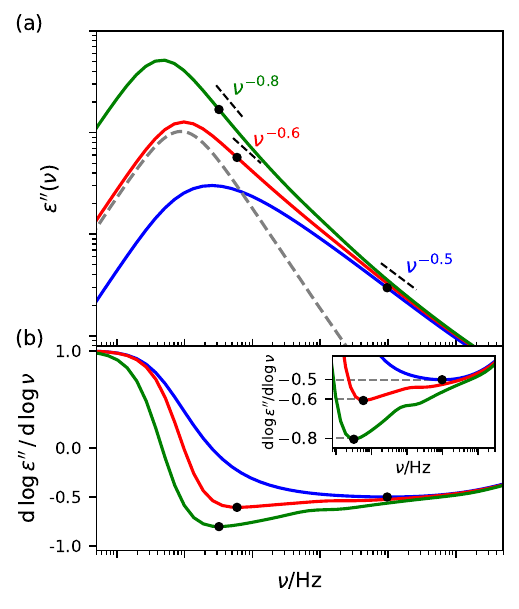}
    \caption{(a) Illustration how the slow Debye-shaped cross-correlation process (dashed line) that superimposes the self-part (blue line) affects the overall shape of the dielectric loss. The blue curve is based on a generalized-gamma distribution of relaxation times with parameters $\alpha=2$ and $\beta=0.5$~\cite{Pabst2021,Blochowicz2003} (b) The exponent of the high frequency power law can be determined in a model-free approach by calculating the minimum value of the the logarithmic derivative, cf. \Eq{equ:beta}}
    \label{fig2}
\end{figure}

Obviously, the physical origin for dipolar cross-correlations must be fairly different among these liquids, e.g., hydrogen bonding, dipole-dipole interactions, $\pi$-$\pi$-interactions and possibly also steric effects might be relevant. Remarkably, the relation between $\gk$ and $\beta$ seems to be identical for the different sub-classes and even includes monohydroxy alcohols, which were excluded from the Paluch correlation~\cite{Paluch2016}. Therefore, the presented results support the qualitative picture that dipolar cross-correlations lead to a steeper high-frequency power law in the dielectric loss of supercooled liquids. 

The fact that the relation between $\gk$ and $\beta$ is the same for different classes of liquids suggests that $\gk$, quantifying \textit{static} dipolar cross-correlations,  also determines \textit{dynamic} dipolar cross-correlations and how they relate to the self-correlations of dipolar molecules.

These results are in good agreement with the results derived recently from the theory of dielectrics by Dejardin et al. In particular, for liquids with $\gk>1$, a distinct slow cross-correlation process is predicted in the dielectric spectrum in addition to the peak reflecting the self-correlations of dipole moments~\cite{Dejardin2019}. While their theory, at least in the current state, does not consider dynamic heterogeneity, one can rationalize what is expected in a scenario when molecular dynamics is heterogeneous following the well-known line of argument by Anderson et al.~\cite{Anderson1967}: In the limit of the cross-correlation process being significantly slower than the self-correlations, the former is also slower than structural relaxation. Therefore, it would contribute to the dielectric loss as a narrow peak associated with only a single average relaxation time, reflecting an average over different heterogeneous environments. 

\begin{figure*}
    \includegraphics[width=0.95\textwidth]{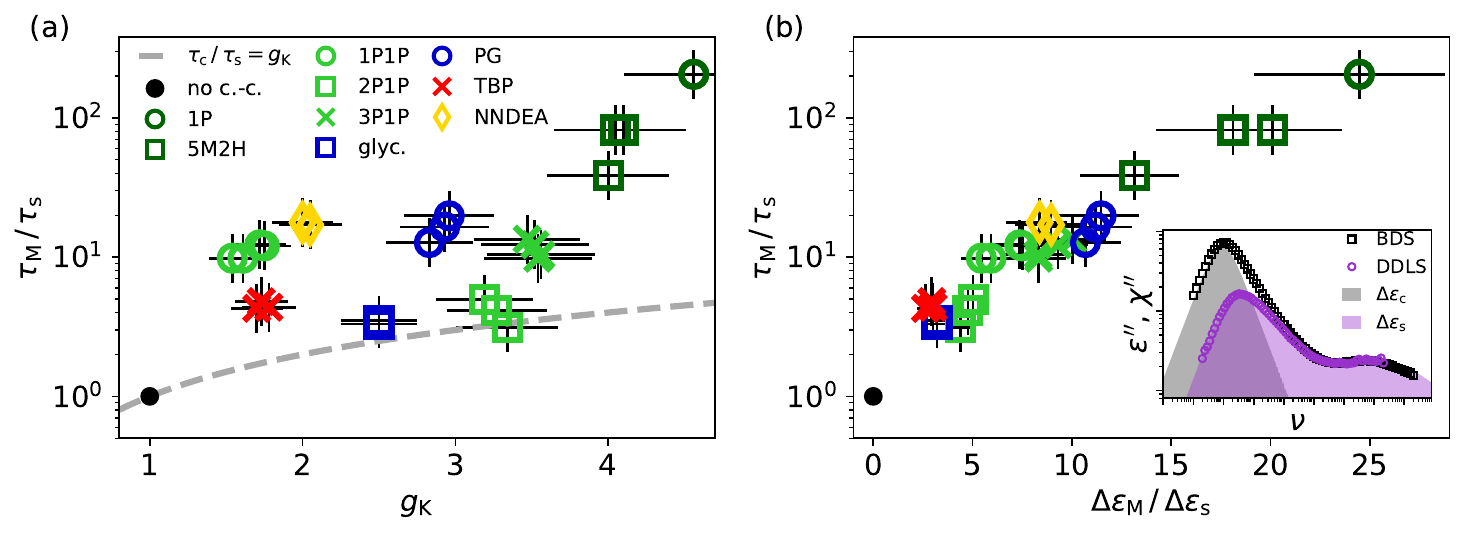}
    \caption{(a) Test of the KM relation (dashed grey line, cf.\ \Eq{equ:KM}) for different supercooled liquids, each studied at several temperatures. (b) Relation between the quotients of relaxation times and relaxation strengths reflecting the collective dielectric response and the self-part of the latter, respectively. The properties of the self-part were determined from a combined analysis of the dielectric loss and DDLS spectra, see the inset. Errorbars for $\tau_\mathrm{M}/\tau_\mathrm{s}$ and $\Delta\varepsilon_\mathrm{M}/\Delta\varepsilon_\mathrm{s}$ reflect the propagation of uncertainty from the assumption $\Delta\tau_\mathrm{s}/\tau_\mathrm{s}=0.5$, to consider, e.g., possible temperature differences between BDS and PCS. In contrast to Fig.\ \ref{fig1} the depicted quantities are not temperature-averaged.}
    \label{fig3}
\end{figure*}

In \fig{fig2}\,a we illustrate how the additional slow and narrow cross-correlation contribution does affect the shape of the dielectric loss spectrum. The blue curve reflects the dielectric loss spectrum found for a supercooled liquid without any cross-correlations ($\gk=1$), thus it is assumed to have $\beta=0.5$ as commonly found in experiments~\cite{Pabst2021,Koerber2020,Becher2021,Becher2021a}. To obtain the red and the green curve, the blue curve is superimposed with an additional slow and Debye shaped process as indicated by the grey dashed line reflecting the slow contribution to the red curve. The resulting high-frequency power-laws are steeper with exponents $\beta=0.6$ and 0.8, respectively. \fig{fig2}\,b demonstrates how $\beta$ is obtained by \Eq{equ:beta}.  Note that red and green curve differ only regarding the relaxation strength of the slow cross-correlation contribution and its dynamic separation with regard to the self-correlation peak. Thus, we find that the high-frequency power law of the combined peak critically depends on the specific properties of dipolar cross-correlations in each liquid, resulting in a variety of different values for $\beta$.

Based on these results we argue that the observations in \fig{fig1} are well in line with the predictions by Dejardin et al.~\cite{Dejardin2019}: The larger $\gk$, the higher the relaxation strength of the additional cross-correlation process, thus the stronger its impact on the dielectric spectrum. As a consequence, for larger $\gk$ the high frequency power law will be more and more dominated by the narrow cross-correlation process compared to the asymmetric self-correlation process. Therefore, $\beta$ is expected to grow with increasing $\gk$, as it is nicely confirmed by the data presented in \fig{fig1} and illustrated in \fig{fig2}. Additionally, the clustering of data points around $\gk\approx 1$ and $\beta=0.5$ supports the hypothesis of a generic shape of the structural relaxation process~\cite{Pabst2021}.

Moreover, the observed correlation between $\gk$ and $\beta$ also rationalizes the origin of the Paluch correlation. In the light of the theory of Dejardin et al.\, the relation between $\gk$ and $\beta$ can be understood as being causal: Large values of $\gk$ are associated with dipolar order, which in turn contributes to the dielectric loss and leads to larger values of $\beta$. At the same time, this implies that the correlation between $\de$ and $\beta$ is more indirect, as it is mediated by $\gk$, with $\gk>1$ leading to larger values of $\de$. Thus, the probability that a liquid with high $\de$ also has $\gk>1$, and therefore $\beta>0.5$, is enhanced compared to liquids with low $\de$.

Having established the correlation between static cross-correlations and the apparent stretching of the dynamics, the question arises how the macroscopic dielectric response and the self correlations of dipolar molecules are related. This question was already addressed many years prior to the theory of Dejardin et al.\, by Keyes\,~\cite{Keyes1972} and later by Kivelson and Madden (KM)~\cite{Kivelson1975,Madden1984}. In their "micro-macro" relations the relaxation time of the collective response, $\tau_\mathrm{M}$ and the relaxation times of single dipoles, $\tau_\mathrm{s}$ are related via
\begin{equation}
    \tau_\mathrm{M} = \gk\,\tau_\mathrm{s}.
    \label{equ:KM}
\end{equation}
Although the validity of the approach by KM was later challenged by Bordewijk\,~\cite{Bordewijk1980}, the KM relations are still discussed in the context of dipolar cross-correlations to date~\cite{Matyushov2023}.

In a final step we aim to test the validity of the KM \Eq{equ:KM} for supercooled liquids. In order to do this we assume that the self-part of molecular dynamics is probed by DDLS, as suggested by prior experimental work\,~\cite{Gabriel2017,Gabriel2018,Boehmer2019,Pabst2020,Pabst2021,Boehmer2022,Boehmer2023} and recent computer-simulation results~\cite{Henot2023}. 
We illustrate the joint analysis of BDS and DDLS spectra in the inset of \fig{fig3}b for the case of tributyl phosphate (cf. Ref.\,~\cite{Pabst2020}). Here, the dielectric loss peaks at a significantly lower frequency and displays a steeper high frequency power law than the DDLS spectrum. As was extensively discussed in previous publications \cite{Pabst2020,Pabst2021}, this discrepancy cannot be solely due to the difference in timescale caused by the different rank Legendre polynomials $\ell = 1, 2$ probed by both techniques. By contrast, and in line with the theory of  Dejardin et al.~\cite{Dejardin2019}, the dielectric loss is described by adding a slow and Debye shaped cross-correlation process to the DDLS spectrum, illustrated by the grey shaded area.  We note that recent MD simulation results support this procedure indicating that cross-correlations dominate the $\ell = 1$ correlation function (dielectric response) in polar substances and are almost negligible in $\ell=2$ functions (light scattering). The self correlations on the other hand turn out to be almost identical.~\cite{Koperwas2022,Paluch2023a,Henot2023} As indicated in the inset of Fig.\ \ref{fig3}b  it is thus straight forward to extract an estimate of the dielectric relaxation strengths $\Delta\varepsilon_\mathrm{s}$ of the self- and $\Delta\varepsilon_\mathrm{M}$ of the collective response and the corresponding peak relaxation times $\tau_\mathrm{s}$ and $\tau_\mathrm{M}$. Note that in contrast to Fig.\ \ref{fig1} averaging of values at different temperatures is not applied.

In Fig.\ \ref{fig3}a we test the KM micro-macro relation by plotting $\tau_\mathrm{M}/\tau_\mathrm{s}$ over $\gk$ for several liquids and at different temperatures. Almost all points deviate significantly from the dashed line that indicates the KM prediction. Instead, the dynamic separation between collective and single-molecule relaxation is larger for most liquids.

In \fig{fig3}b we instead find a common relation between $\tau_\mathrm{M}/\tau_\mathrm{s}$ and $\Delta\varepsilon_\mathrm{M}/\Delta\varepsilon_\mathrm{s}$, as in this representation data for all investigated liquids collapse onto a single curve. Such a trend was previously reported for monohydroxy alcohols and rationalized in terms of the transient chain model~\cite{Boehmer2014,Gainaru2010}, as both,  $\tau_\mathrm{M}/\tau_\mathrm{s}$ and $\Delta\varepsilon_\mathrm{M}/\Delta\varepsilon_\mathrm{s}$, are expected to increase with increasing length of hydrogen bonded supra-molecular chain-structures. The interesting conclusion from \fig{fig3}b is that also polyhydric alcohols and liquids that interact via dipole-dipole interactions display the same relation and lie on the same common line in the respective plot. This suggests the relation between $\tau_\mathrm{M}/\tau_\mathrm{s}$ and $\Delta\varepsilon_\mathrm{M}/\Delta\varepsilon_\mathrm{s}$ to be of much more fundamental nature and to represent a common characteristic shared by dipolar cross-correlations of diverse origin. In its essence, the finding from \fig{fig3}b is similar to the identification of a spectral envelope for dielectric loss spectra reported by Gainaru~\cite{Gainaru2019}. Our results now suggest that the physical origin of the identified envelope likely is related to the somewhat universal way how cross-correlations contribute to the dielectric loss. 

To summarize, we have determined the Kirkwood correlation factor $\gk$ and the high-frequency power law exponent $\beta$ for 25 different supercooled liquids in order to answer the question how dipolar cross-correlations affect the shape of the dielectric loss spectrum. Our analysis reveals that large values of $\beta$ are always associated with large values of $\gk$. Once dipolar cross-correlations are absent ($\gk=1$), we find $\beta\approx0.5$ in line with recent experimental results from DDLS and NMR. We discuss our results with regard to the recent theory of Dejardin at al.\ that predicts an additional slow cross-correlation process in the dielectric loss for liquids with $\gk>1$~\cite{Dejardin2019}. It is shown that based on this result, the high-frequency power law of the dielectric loss is expected to be steeper for liquids with strong dipolar cross-correlations, as it is confirmed by the experimental data.

Under the assumption that DDLS probes the self-correlation part of molecular dynamics we test the "micro-macro" relation by Kivelson and Madden and find that it is strongly violated by most of the  studied supercooled liquids. Instead we find a common relation between the ratio of time constants and relaxation strengths of the collective dielectric response and its self-part, which is obeyed by supercooled liquids with very different molecular structure.

\begin{acknowledgments}
The authors thank Pierre-Michel Déjardin and Rolf Zeißler for fruitful discussion, Ranko Richert for providing the m-toluidine data, Marian Paluch for providing the salol data and Ernst Rößler for providing the toluene data. Financial support by the Deutsche Forschungsgemeinschaft under grant no. BL 1192/3 and the VILLUM Foundation’s Matter grant (16515) is gratefully acknowledged.
\end{acknowledgments}

\bibliography{gk}

\end{document}